\documentclass{article}
\usepackage{graphicx}
\usepackage{cite}
\usepackage{amssymb}
\usepackage[utf8]{inputenc}

\usepackage[T1]{fontenc}

\def\mhyph{{\hbox{-}}}

\begin{document}
\title{RDM-stars as sources of fast radio bursts}
\author{Igor Nikitin\\
Fraunhofer Institute for Algorithms and Scientific Computing\\
Schloss Birlinghoven, 53757 Sankt Augustin, Germany\\
\\
igor.nikitin@scai.fraunhofer.de
}
\date{}
\maketitle

\begin{abstract}
In this work we consider a recently formulated model of a black hole coupled with radially directed flows of dark matter (RDM-star). In this model, a cutoff by quantum gravity (QG) creates a core of Planck density, filled with a gas of Planck mass particles, inside the star. 

Further, in this model, a fast radio burst (FRB) can be generated via the following mechanism. An object of an asteroid mass falls onto an RDM-star. Due to large gravitational forces available in the interior, the RDM-star works as a powerful accelerator, boosting the nucleons composing the object to extremely high energies. Upon collision with the core, the nucleons enter in deeply inelastic reactions with the Planck particles composing the core. This process is followed by stimulated emission of highly energetic photons. On the way outside, these photons are subjected to a strong redshift factor, downscaling their frequencies to the radio diapason, as a result producing an FRB. 

A straightforward computation gives a characteristic upper FRB frequency of 0.6GHz. With attenuation factors, it is stretched to a range 0.35-8GHz, fitting well with the observable FRB frequencies. We also discuss the reconstruction of other FRB parameters in frames of the model.
\end{abstract}

\section{Introduction}

Fast radio bursts (FRBs) are superpowerful flashes observed in the radio band, whose nature is a puzzle to astrophysicists. The lowest frequency at 111 MHz was registered on the BSA LPI telescope in Pushchino, 100 km south from Moscow, for the bursts FRB151018, FRB160920, FRB170606 (YYMMDD). They were found in postprocessing of archived data and reported on 10-aug-2018 \cite {ATel11932}. The next up in frequency at 580 MHz, FRB180725A was detected in the Canadian project CHIME \cite {ATel11901}. It is followed by a large cluster of observations near 1GHz, containing also the first detected burst FRB010724, by Lorimer et al. 2007 \cite {07094301}. The bursts are cataloged by FRBCAT \cite {frbcat}, enlisting, at the time of writing this paper, 66 FRB sources. Of these, only for one, FRB121102, many repetitions were recorded \cite {160300581,160308880}. The upper frequency at 8GHz was recorded on the GBT telescope, USA, for the repeater \cite {180404101,180903043,181110748}. The duration of the bursts varies from fractions of milliseconds at high frequencies to seconds at low frequencies. The estimates of the distances to the sources by dispersion shift indicate their extragalactic origin. The total isotropic energy of the burst is of the order $ \sim10 ^ {32 \mhyph34} $~J, corresponding to the rest energy $ mc ^ 2 $ for the mass of a moderate size asteroid.

To describe the nature of FRB, many hypotheses have been proposed. They have been connected to gamma-ray bursts \cite {13104893}, hyperflares of magnetars \cite {07102006}, collapse of pulsars \cite {13071409,14126119}, superradiance \cite {171000401}, cosmic masers \cite {170306723}, axion miniclusters \cite {14113900}, cosmic strings \cite {cosmstr}, even to extraterrestrial intelligence \cite {170101109}. An FRB hypothesis based on quantum gravity (QG) was put forward in \cite {14094031}. This paper considered {\it Planck stars}, objects also studied in \cite {14016562,151100633} and references therein. According to this model, the collapse of matter into a black hole, upon reaching densities of the order of Planck one, stops and is changed to a quantum bounce. Such a phenomenon is also present in cosmological models \cite {bigbounce}, in which the Big Bang is preceded not by a mathematical singularity, but by the Big Bounce event, before which the universe underwent a stage of contraction. In the Planck stars model, this phenomenon occurs in a miniature, the matter that fell into a black hole flies back in a T-symmetric manner, thereby effectively making the black hole white. This process requires a very short time in the reference frame of falling matter, $ \sim2r / c $, where $ r $ is the size of the object, $ c $ is the speed of light, from picoseconds for micro-holes $ r \sim10 ^ {- 4 } $ m to minutes for supermassive black holes $ r \sim10 ^ {10} $ m. However, the superstrong redshift existing inside the object slows down this process, as seen by a remote observer, to times of the order of the age of the universe. The corresponding bounce time depends on the size of the object, and, according to the calculation \cite {14094031}, primordial micro-holes of size $ r \sim2 \cdot10 ^ {- 4} $ m are breaking up now. This should lead to a characteristic wavelength of $ \lambda \sim r $, corresponding to a frequency of $ 1.5 $ THz, which is 3 orders of magnitude higher than the typical frequencies of FRB observations, and a total burst energy of $ 10 ^ {40} $ J, that 6 orders of magnitude higher than the isotropic energies of the observed FRB. Note that the difference by several orders of magnitude is normal for the astrophysical models based on QG, where very different scales of energy, distance and time appear, forming dimensionless combinations, such as the ratio of the age of the universe to the Planck time $ t_H / t_P \sim8 \cdot10 ^ {60} $. A previous estimate for the signal wavelength, corresponding to the Hawking evaporation of primordial black holes \cite {14016562}, a slower QG process than the bounce of Planck stars, is $ 10 ^ {- 16} $ m, gamma rays.

In this paper, we will consider a variant of dark stars, coupled to radial dark matter flows (RDM), a model that we formulated in \cite {static-rdm}. Later, in the work \cite {wrmh-rdm}, the possibility of opening {\it a wormhole} in the center of the model was investigated, and in the work \cite {wh-stability} a related question of so called Eardley's instability \cite {Eardley, OriPoisson} of {\it white holes}.

The RDM model has some properties similar to Planck stars, namely, the presence in the depth of these stars of a very large, almost arbitrary redshift factor, mathematically limited by numbers $ \sim\exp (10 ^ 6) $. For Planck stars, the redshift factor is limited by QG-cutoff, which we also impose in our model. The difference from Planck stars is that the RDM model is stationary and is not related to cosmological processes. It represents a kind of Planck star, in which evolution stopped and came to an equilibrium with the pressure of the dark matter flows surrounding the star. To generate an FRB in this model, an external trigger is needed, for which we will choose an asteroid falling onto RDM-star. In Section 2, we present the typical solution for the RDM-star, inside which the Planck core was formed, as a result of the QG-cutoff performed in the model. There, we will also estimate the frequency range for the FRB produced by RDM-star when external objects fall on it. In Section 3, we discuss other FRB parameters in frames of RDM model, as well as constraints on the RDM model imposed by observable FRB characteristics. In conclusion, we briefly summarize the results.

We use the geometric unit system $G=c=1$ throughout this paper.

\section{RDM-star and QG-cutoff}

\begin{figure}
\begin{center}
\includegraphics[width=\textwidth]{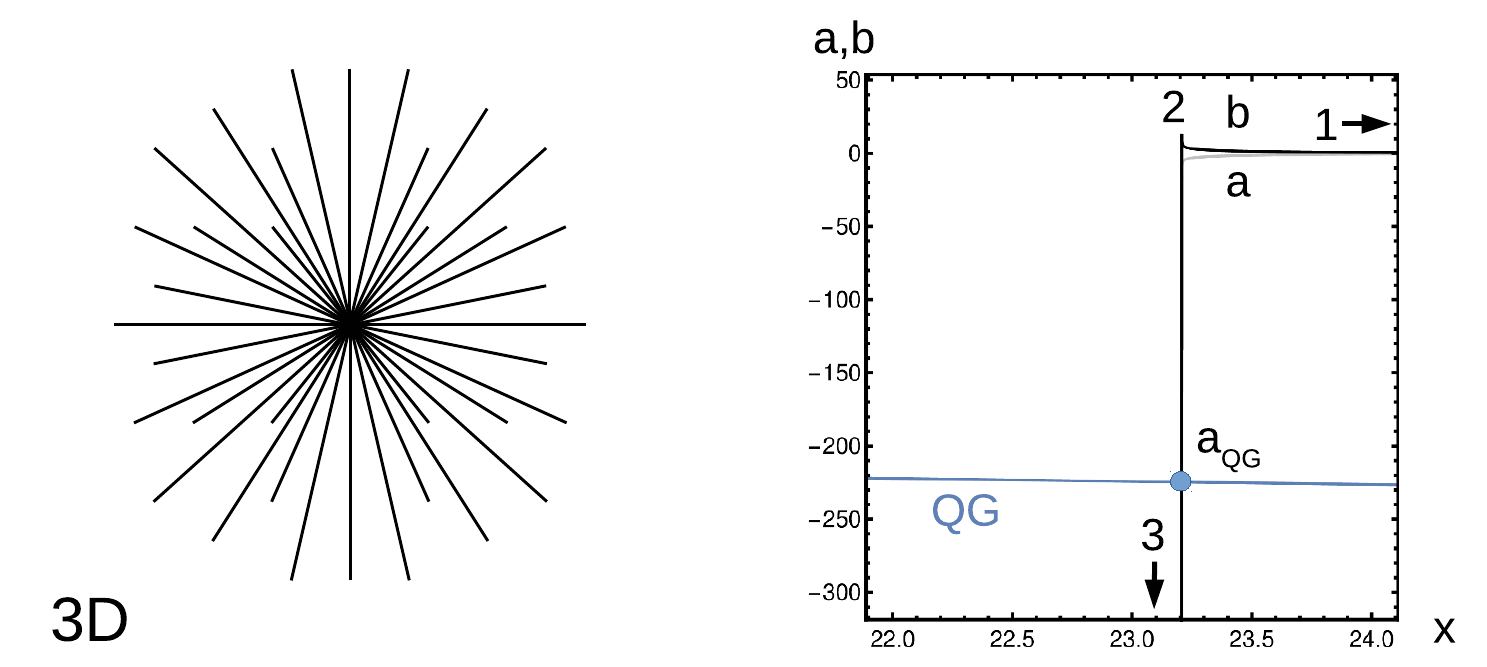}
\end{center}
\caption{On the left: RDM-star, a black hole coupled to spherically symmetric radial dark matter flows. On the right: behavior of metric coefficients, for RDM-star with parameters of Milky Way galaxy. QG-line shows the quantum gravity cutoff.}\label{f1}
\end{figure}

The RDM model has been formulated in our works \cite {static-rdm, wrmh-rdm, wh-stability}. It considers spherically symmetric radial dark matter flows, shown in Fig.\ref {f1} on the left. Each fiber represents a T-symmetric superposition of two flows, converging towards the center and diverging from the center. In this configuration, the energy flows through the spheres surrounding the center compensate each other, the mass bounded by the spheres is conserved, the solution does not depend on time, is stationary.

An interesting feature of this model is that the orbital velocities of objects rotating around the center, at a sufficiently large distance from it, do not depend on the distance to the center. In other words, the model has flat rotation curves. Indeed, the geometry of the system in the form of diverging dark matter fibers implies a density dependence, inversely proportional to the square of the distance, $ \rho_ {grav} = \epsilon / (4 \pi r ^ 2) $, in geometric units, where $ \epsilon $ is a constant coefficient of proportionality. From here we obtain a mass function linearly increasing with distance, $ M_ {grav} = \epsilon r $, orbital accelerations in Newtonian approximation $ v ^ 2 / r = M_ {grav} / r ^ 2 $, whence the square of the orbital velocity is constant and is equal to $ v ^ 2 = \epsilon $. Because of this property, the RDM configuration can be used as a model of dark matter for spiral galaxies, for which, as it is well known, the experimentally measured rotation curves at large distances are also approximately constant \cite {SofueRubin}. Substituting here experimentally measured velocities for the Milky Way (MW) galaxy: $ v \sim200 $ km/s, we get in geometrical units $ v / c \sim6.6 \cdot10 ^ {- 4} $ and $ \epsilon \sim4 \cdot10 ^ {- 7} $.

The solution of Einstein field equations (EFE) describing the behavior of the RDM model in strong gravitational fields was carried out in \cite {static-rdm, wrmh-rdm, wh-stability} using numerical methods. The parameters of the solution for the MW galaxy were given in tabular form in these papers, while here we show a piece of the solution in the graph Fig.\ref {f1} on the right.

The solution is given for the metric coefficients $ A, B $, where the first coefficient determines the effect of time dilation $ dt_ {in} = dt_ {out} A ^ {1/2} $, and the second one -- the effect of the curvature of the space $ dL_ {rad} = dr B ^ {1/2} $. Here $ dt_ {in, out} $ is the time measured by an observer at rest, respectively, inside the object and outside, at a large distance from the object, while the value of the $ A $-function at large distances is set to 1 by an agreement. The same function defines the redshift of the wavelength $ \lambda_ {out} = \lambda_ {in} A ^ {- 1/2} $ for the light coming out of the object, as well as the inverse effect of blueshift and, in general, an increase in the energy of any kind of matter falling on the object, $ E_ {in} = E_ {out} A ^ {- 1/2} $. Another metric coefficient determines the variation of the proper element of length in the radial direction $ dL_ {rad} $ with respect to the variation of the radial coordinate $ dr $. This radial coordinate, also called {\it aerial radius}, defines an element of proper length in the tangential direction, by the standard formula for the variation of the angle $ dL_ {tan} = rd \theta $, as well as the area of $ r $-sphere as $ 4 \pi r ^ 2 $. Further details about EFE and metric coefficients can be found in the Appendix.

Due to the strong variation of the metric coefficients on the RDM solution, it is given in logarithmic coordinates:
\begin{eqnarray}
&& x=\log r,\ a=\log A,\ b=\log B. \label{xab}
\end{eqnarray}
Even taking the logarithm here is not enough, the resulting solution is difficult to depict on one graph. In Fig.\ref {f1} we show the part of the solution that interests us. The starting point of integration 1 corresponds to the edge of the MW galaxy and is located outside the graph, with $ x_1 \sim49.5 $. Further, the solution reaches a maximum of $ b $ at point 2, while the other coefficient falls approximately symmetrically as $ a \sim-b $. This behavior corresponds to Schwarzschild regime, for which the functions $ A, B $ are inverse to each other, $ A = 1-r_s / r $, $ B = A ^ {- 1} $, where $ r_s $ is the Schwarzschild radius. The values of $ B $ reached at point 2 are really large $ B_2 \sim6.25 \cdot10 ^ 5 $. Next comes the mass inflation mode, in which $ a, b $ fall almost parallel to each other, $ b = a + Const $, with almost constant $ x $. Further, the solution reaches a minimum of $ a $ at point 3, which is located outside the graph and will not interest us.

Our goal now is to impose the appropriate QG-cutoff on the solution. Following \cite {14016562}, the QG-cutoff criterion is imposed when the solution reaches Planck density, in geometric units $ \rho \sim l_P ^ {- 2} $, where $ l_P $ is the Planck length. This is a perfectly reasonable requirement, since according to EFE, the space-time curvature is proportional to the density, and the appearance of QG effects is natural when the curvature reaches Planck values. In \cite {14016562}, it was also noted that this condition can be fulfilled much earlier, before the geometric dimensions of the solution reach the Planck scale. We will now impose this condition in the RDM model.

For definiteness, we will consider null radial dark matter (NRDM) \cite {wh-stability}, although the solution structure weakly depends on the type of matter. The density and pressure of matter is determined by the formula
\begin{eqnarray}
&&\rho_{\mbox{\it\footnotesize eff}}=p_{\mbox{\it\footnotesize eff}}=\epsilon/(8\pi r^2A).\label{rhopeff}
\end{eqnarray}
Note that effective gravitating density at large distances is determined by the sum $ \rho_ {grav} = \rho _ {\mbox {\it \footnotesize eff}} + p _ {\mbox {\it \footnotesize eff}} \sim \epsilon / ( 4 \pi r ^ 2) $.

The condition $ \rho _ {\mbox {\it \footnotesize eff}} \sim l_P ^ {- 2} $ is reached in the region of a rapid drop of the coefficients $ a, b $, corresponding to the phenomenon of mass inflation. This condition cuts off most of the interesting structures of the solution. Since the fall of the coefficients occurs at almost constant $ r \sim r_s $, the achieved values of $ A $ have the form
\begin{eqnarray}
&&A_{QG}=\epsilon\,(l_P/r_s)^2/(8\pi).\label{Aqg}
\end{eqnarray}
The numerical values of this coefficient for the parameters of MW galaxy correspond to the redshift factor $ A_ {QG} ^ {1/2} = 1.7 \cdot10 ^ {- 49} $.

Thus, we have obtained a key formula that determines the redshift factor for the photons emitted from the surface of the Planck core, as well as an increase in the kinetic energy of the matter falling from the outside onto the Planck core. How the Planck core is composed is unknown. We will simply assume that the core is filled with a fluid or a gas consisting of particles of a mass of the order of Planck one.

Further, for the formation of FRB, we propose the following mechanism. The characteristic total burst energy \cite {0193} reaches $ \sim10 ^ {32 \mhyph34} $ J, or $ \sim10 ^ {15 \mhyph17} $ kg in mass equivalent, which is the mass of an asteroid. First, we assume an energy conversion efficiency of the order of unity and consider an external object of the order of the mass of an asteroid falling on an RDM star. As a result of the action of tidal forces and superstrong acceleration, the object will be split into component particles, for which we will consider nucleons, that is, protons or neutrons. In fact, the splitting can go further, to quarks, or stop earlier, at the level of nuclei. The RDM star acts as a superstrong particle accelerator, boosting the particles falling on it to ultrarelativistic velocities corresponding to the particles energy $ E_N \sim m_N A_ {QG} ^ {- 1/2} $, where $ m_N $ is their rest mass. These accelerated particles fall onto the Planck core and are scattered on it as on a fixed target. Accelerated particles of energy $ E_N $ enter into reactions with the particles of mass $ m_X $ composing the core. In the reactions, high-energy photons are released that go outside with the redshift factor $ A_ {QG} ^ {1/2} $. These photons represent FRB.

Further estimations depend on the type of reactions occurring inside the Planck core. First, we assume that as a result of the reactions, photons are born with the energy of the order of the energy of the accelerated particles $ E _ {\gamma, in} \sim E_N $, as it occurs, for example, for {\it Bremsstrahlung} near the cutoff frequency. Since the acceleration and the redshift factors cancel each other, the outgoing photons have energy $ E _ {\gamma, out} \sim m_N $ in the region of 1~GeV, the wavelength is of the order of the Compton length for the nucleon $ \sim10 ^ {- 15} $ m, gamma radiation. Interestingly, this answer is close to the estimate $ \sim10 ^ {- 16} $ m \cite {14016562} for Planck stars, obtained from completely different considerations.

Since the typical FRB wavelengths differ by 14 orders of magnitude from this value, it is necessary to look for another mechanism. Suppose that inelastic scattering of incoming particles on $X$-particles occurs, in which an excited state $ X_1 ^ * $ is born with the energy $ E (X_1 ^ *) \sim \sqrt {2 m_X E_N} $. This is the maximum amount of energy that can be released in the inelastic collision of $ m_N $ and $ m_X $ in their center of mass system, which follows from the kinematics of the process at $ E_N \gg m_X \gg m_N $. After this collision, the excited particle and the nucleon move with the same speeds, and due to $ E (X_1 ^ *) \ll E_N $, their energy is almost equal to the original $ \sim E_N $. Further, these particles enter into the inelastic collision with the next $X$-particle, which also goes into the excited state, $ X_2 ^ * $. This {\it snowball effect} continues until all the kinetic energy of the initial particle is spent. Thus, a column of excited particles will be born, each of energy $ E (X_n ^ *) \sim E (X_1 ^ *) n ^ {- 1/2} / 2 $, which will be confirmed by the calculation below. The specific form of this spectrum is currently not so important, note that the energy of the first excited state $ E (X_1 ^ *) $ scales the entire spectrum and cuts it from above. At the end of this process, all particles will be stopped, and all the initial kinetic energy will be deposited into the energy of the excited states $ X_n ^ * $.

Then the particles transfer from the excited state to the initial state with the emission of one photon: $ X_n ^ * \to X + \gamma $. Due to the fact that $ E (X_n ^ *) \gg m_X $, the photon receives only a half of the excitation energy of the particle, $ E _ {\gamma, in} \sim E (X_n ^ *) / 2 $. The other half goes into recoil of the particle $X$, and, ultimately, into the thermal energy. Already here we see that only a part of the initial kinetic energy of the asteroid can be transformed into FRB energy, and the energy relations will require corrections, which we will consider below.

The observed energies of the produced photons, at the maximum frequency, after taking into account the redshift are $ E _ {\gamma, out} = E (X_1 ^ *) / 2 \cdot A_ {QG} ^ {1/2} $, which, after substitutions, gives the wavelength 
\begin{eqnarray}
&&\lambda_{out}=\sqrt{2\lambda_X \lambda_N}\,A_{QG}^{-1/4},\label{lamout1}
\end{eqnarray}
where $\lambda_X$ and $\lambda_N$ are Compton wavelengths of particles $X$ and $N$ respectively. Assuming $\lambda_X=l_P$, which corresponds to the mass $m_X=2\pi m_P$, and substituting (\ref{Aqg}), obtain
\begin{eqnarray}
&&\lambda_{out}=2 (2\pi)^{1/4} \sqrt{r_s\lambda_N}\,/\epsilon^{1/4}.\label{lamout2}
\end{eqnarray}
Substituting here the Compton length of the nucleon and galactic parameters \cite{SofueRubin,Ghez}: 
\begin{eqnarray}
&&\lambda_N=1.32\cdot10^{-15}\textrm{m},\ r_s=1.2\cdot10^{10}\textrm{m},\ \epsilon=4\cdot10^{-7}, \label{lampar1}
\end{eqnarray}
obtain
\begin{eqnarray}
&&\lambda_{out}=0.5\textrm{m},\ \nu_{out}=0.6\textrm{GHz},\label{lamout3}
\end{eqnarray}
that falls in 0.111 ... 8GHz, the characteristic range of FRB observations, exactly to the central frequency of CHIME/FRB180725A observation.

In fact, due to many side effects, it is not currently possible to make an accurate calculation of the observed FRB characteristics. These signals are broadband, a single observation usually covers a range of the frequencies. The repeated signal FRB 121102 was recorded on 111MHz \cite {ATel11932}, it was also active around 1GHz \cite {160300581,160308880}, and its maximum observed frequency is 8GHz \cite {180404101,180903043,181110748}. This suggests that the frequency of this FRB source may float or have components in a wide band. The observed wavelength is also subject to a cosmological redshift, in the range of $z=0.2$-$1.5$ \cite {0193}. At the same time, the value of $ z $ is not precisely measured, but inferred with a number of additional assumptions, discussed in more detail below. If we treat this correction as a measurement uncertainty, the frequency emitted on the host can be up to $ 1 + z \sim2.5 $ times higher than the observed one.

Also, our simple model will only make an estimate in order of magnitude. This estimate represents the upper bound for the observed FRB frequencies corresponding to the maximum possible energy of the emitted photons in the considered type of reactions. This estimate is valid only in order of magnitude, since the requirement that the solution reaches the Planck density really does not give a sharp switch of QG-effects, but only an approximate position. One can introduce an attenuation factor, which marks the switch of QG-effects with a margin, when the density is $ \rho = l_P ^ {- 2} / s_1 $. The final result depends mildly on this factor and we will allow a guess $ s_1 = 1 ... 10 $ for its value. Also, we noted that the original particles can be not nucleons, but quarks or nuclei, which gives another factor $ \lambda_N \to \lambda_N / s_2 $, varying between $ 1/3 $ for the constituent quarks and $ \sim56 $ for iron nuclei. With these corrections we get
\begin{eqnarray}
&A=s_1\epsilon\,(l_P/r_s)^2/(8\pi),\label{Aqg2}\\
&\lambda_{out}=s_1^{-1/4}s_2^{-1/2} 2 (2\pi)^{1/4} \sqrt{r_s\lambda_N}\,/\epsilon^{1/4},\label{lamout4}\\
&\lambda_{out}=0.038...0.87\textrm{m},\ \nu_{out}=0.35...8\textrm{GHz},
\end{eqnarray}
which gives a good overlap of the observed frequency band by our prediction.

\section{Discussion}

The frequency range of the FRB is not the only, but the most stably predictable feature. Let's discuss the possibility of reconstructing other parameters.

{\it Energy spectrum of the burst.} Let us determine the frequency distribution of energy for the snowball effect described above. Considering the compound $ Y_n $ formed by the original $ N $-particle and the excited $ X_n ^ * $-particles sticking to it, and using the energy-momentum conservation laws, we get $ E (Y_n) = E_N + nm_X $, $ P (Y_n) = P_N $, $ m (Y_n) = ((E_N + nm_X) ^ 2-P_N ^ 2) ^ {1/2} = (m_N ^ 2 + 2nm_XE_N + (nm_X) ^ 2) ^ {1/2 } $, where $ E, P, m $ denote energy, momentum and mass of the considered particles. The mass of the excited state is $ m (X_n ^ *) = m (Y_n) -m (Y_ {n-1}) $. We consider the limit $ E_N \gg m_X \gg m_N $. For the first state, $ m (X_1 ^ *) \sim (2m_XE_N) ^ {1/2} $. Further, for $ 1 \ll n \ll 2E_N / m_X $ in the expression for $ m (Y_n) $, the term $ 2nm_XE_N $ prevails under the root, as a result we get $ m (X_n ^ *) \sim m (X_1 ^ *) n ^ {- 1/2} / 2 $. The excitation energy that can be converted into photons is $ E (X_n ^ *) = m (X_n ^ *) - m (X) $, in the considered limit $ E (X_n ^ *) \sim m (X_n ^ *) $.

When $ n \gg 2E_N / m_X $, the term $ (nm_X) ^ 2 $ prevails in $ m (Y_n) $, $ m (X_n ^ *) \sim m_X $ results, there are practically no excitations, the snowball is stopped. Due to $ m_X \ll m (X_1 ^ *) $, as well as after accounting for superstrong redshifts, this limit is realized in the low-energy part of the spectrum outside the FRB detection range.

Thus, in the spectral region of interest immediately adjacent to the threshold $ E (X_1 ^ *) = (2m_XE_N) ^ {1/2} $, $ E (X_n ^ *) \sim E (X_1 ^ *) n ^ {-1/2} / 2 $. The number of photons in a given frequency interval is proportional to $ | dn / dE | $, while the energy density is proportional to $ E | dn / dE | = | d \log E / dn | ^ {- 1} \sim 2n \sim (2E / E (X_1 ^ *)) ^ {- 2} $, which implies the initial energy of the process with the spectral index $ \alpha = -2 $, $ E | dn / dE | \sim E ^ \alpha $. The redshift scales the common factor $ E (X_1 ^ *) $ and does not change the density $ E | dn / dE | $ and its spectral index.

Note also that the spectral index is usually measured for flux density, and is also divided by the duration of the burst in a given frequency band. The process considered here corresponds to the initial flux density with a spectral index of $ -2 $, assuming that the intrinsic burst duration is constant in frequency.

The spectral index calculated here corresponds to a specific mechanism, the snowball effect, used to generate the excited states. This index may change when considering other mechanisms, such as particle showers, collision cascades, the scattering of high-energy nucleons on gas of Planck particles, as well as the scattering of the produced photons, both on the Planck gas and on the surrounding dark star environment. Experimental measurements give a spectral index that varies widely, $ -10.4 ... 13.6 $, according to FRBCAT. This may mean high variability of this parameter between different bursts, as well as the inaccuracy of its experimental determination. The cutoff frequency is a more stable parameter, since it is determined by the maximum energy that can be deposited and then radiated as photons, and it is characteristic of scattering of high-energy particles on a fixed target.

{\it Total energy of the burst} is determined by the mass of the object falling on the RDM star. According to the mechanism discussed above, half of the rest energy of an object is converted into photons, with associated recoil efficiency $ \eta_ {rec} = 0.5 $. At the same time, the energy spectrum for the considered process is strongly skewed to the left. If we fix $ n = 100 $ and limit FRB observations to the frequency range $ [\nu_ {out} / 10, \nu_ {out}] $, the radiation energy in this region will be determined by $ m (Y_n) = (2nm_XE_N) ^ {1/2} = 10E (X_1 ^ *) $, which is a fraction of the total available energy $ \eta_ {spec} = 10E (X_1 ^ *) / E_N = 10 (2m_X / m_N) ^ {1/2} A_ {QG} ^ {1/4} = 10 (2 \lambda_N / r_s) ^ {1/2} (\epsilon / (8 \pi)) ^ {1/4} $. The numerical value of the spectral efficiency is $ \eta_ {spec} \sim5 \cdot10 ^ {- 14} $. To compensate for this factor, the mass of the falling object must be $ 2 \cdot10 ^ {13} $ times larger, i.e., $ \sim2 \cdot10 ^ {28 \mhyph30} $ kg, from 0.01 to 1 solar mass.

On the other hand, the reported experimental estimates for the total energy use the assumption that the burst is isotropic, which may also require corrections if the radiation has the form of a narrow beam. In this case, the total burst energy and the required mass of the external object can be significantly less than the isotropic estimate. Considering an object of diameter $ D $ on the surface of a sphere of radius $ r_s $, the angular size of the beam can be $ D / r_s $. As we showed in Appendix, a peculiar phenomenon of {\it gravitational beaming} occurs in the depth of the RDM-star, as a result of which the emission of photons in strictly radial direction becomes preferable, this supports the beam geometry considered here. On the other hand, after leaving the RDM star, the wavelength of photons increases and the beam broadening starts to dominate according to the wave optics law $ \lambda / D $. The efficiency of the beam energy transfer compared to isotropic one is $ \eta_ {beam} = 4 \pi / ((\pi / 4) (\lambda / D) ^ 2) = 16D ^ 2 / \lambda ^ 2 $, for $ D = 10 $ km and the considered frequency range $ \sim10 ^ {8 \mhyph12} $. This compensates a significant part of the estimated spectral losses. Thus, a falling object can still be a mass of the order of an asteroid, if the energy leakage in the low-frequency region is compensated by the radiation in the form of a narrow beam.

One can make this estimate somewhat more accurate, assuming that $ D_0 = 10 $~km, $ m_0 = 10 ^ {15} $ kg is used as the default parameters for the falling object. According to the NASA Asteroid Fact Sheet \cite {nasa}, in the Solar System there are asteroids of size $ 0.1 ... 1000 $ km and masses $ 10 ^ {10 \mhyph21} $ kg. Consider the other value of the diameter of $ D $, assuming that the asteroid has the same density. The rest energy will increase by the factor $ (D / D_0) ^ 3 $. Requiring that
\begin{equation}
(D/D_0)^3 \,\eta_{rec}\,\eta_{spec}\,\eta_{beam}=1,\label{etas}
\end{equation}
we get $ D = 20 ... 110 $ km, an asteroid larger than the original $ D_0 $, for which an FRB with an isotropic energy $ m_0 $ will be observed, within the frames of the losses modeled here.

Note that these estimates are highly model dependent. The low-frequency components of the spectrum take up most of the energy and can be completely lost in noise due to the large scatter broadening discussed below, as well as other effects. In this case, the low value of $ \eta_ {spec} $ is justified. However, other models may have a different spectral index, $ E | dn / dE | \sim E ^ \alpha $. For example, for $ \alpha> 0 $, the spectrum is skewed to the higher frequencies, and the leakage to low frequencies can be neglected. The integral of this dependence between the frequencies $ [\nu_ {min}, \nu_ {max}] $ is proportional to $ (\nu_ {max} ^ {\alpha + 1} - \nu_ {min} ^ {\alpha + 1}) / (\alpha + 1) $, and already at $ \alpha> -1 $ the contribution of low frequencies to the integral can be neglected. For such models $ \eta_ {spec} \sim1 $. On the other hand, the assumption of radiation in the form of a narrow beam may also not be fulfilled if the object incident on the Planck core before goes through the phase of splitting and smearing in orbits around the RDM star. In this case, it can also be $ \eta_ {beam} \sim1 $. Our estimations should be understood in the sense of {\it feasibility}, checking the possibility of generating FRB with the observed characteristics when objects with given masses and sizes fall on it. The result is that FRB in the RDM model can be triggered by an asteroid.

\begin{figure}
\begin{center}
\includegraphics[width=0.8\textwidth]{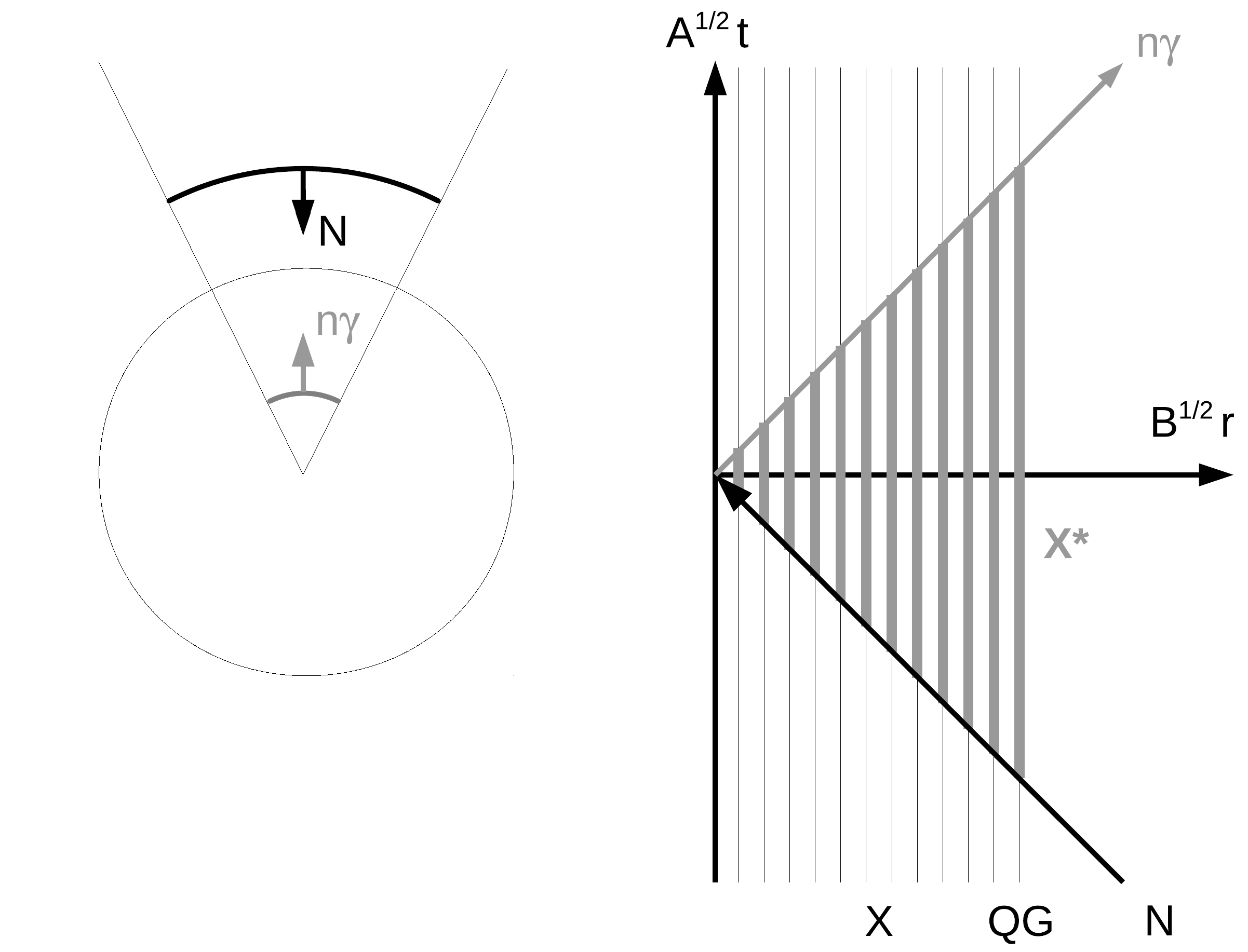}
\end{center}
\caption{A mechanism of coherent FRB emission in the RDM model. On the left: asteroid transformed to a Lorentz-contracted ultrarelativistic thin shell ($N$) falls onto the Planck core and produces a shockwave, energizing the medium in the core. Then the shockwave bounces from the center and stimulates an emission of photons ($n\gamma$) from the energized medium. On the right: temporal diagram of the process.}\label{f2}
\end{figure}

{\it Coherence of the radiation} is usually explained by superradiance \cite {171000401} or stimulated emission \cite {170306723}. We consider here the more general mechanism, stimulated emission, which is the basis of the work of lasers or masers. This mechanism is illustrated in Fig.\ref {f2}. At the top left, an asteroid is shown, transformed to a thin ultrarelativistic shell by Lorentz contraction. Due to gravitational beaming, radial trajectories are preferred for both incoming nucleons $ N $ and outgoing photons $ \gamma $. Upon striking the surface of the Planck core, the shell excites a shockwave inside the conical area of the core marked in the figure. The Planck particles in this zone are excited $ X \to X ^ * $, as a result of the simple snowball effect that we used for the estimates, or another mechanism that has the same inelastic scattering threshold $ E_N $ on $ m_X $. The shockwave propagates close to the speed of light and reaches the center. Next comes the bounce of the shockwave. The mechanism of this bounce is the same as that of Planck stars, the compression of matter above the Planck density leads to the effect of gravitational repulsion; we will consider this effect in more detail below. Further, the diverging shockwave initiates the reverse transitions $ X ^ * \to X + \gamma $, stimulating the emission of photons in the conical region and causing them to coherently release all the stored energy.

The figure on the right shows the temporal diagram of this process. For definiteness, we consider the metric coefficients frozen, constant inside the Planck core. The incoming shell ($ N $) and the shockwave it initiates cause $ X \to X ^ * $ transitions, the excited states formed are shown in the figure by vertical thick gray lines. Further, the bounced shockwave causes the reverse transitions $ X ^ * \to X + \gamma $, the resulting photons multiply each other ($ n \gamma $) and move synchronously with the outgoing shockwave.

{\it Temporal characteristics of the burst.} The observed burst duration cannot be predicted within our simple model. For processes based on stimulated emission, one spontaneously emitted photon in its path causes the emission of subsequent photons synchronous with it, which can give very short durations. Under laboratory conditions, ultrashort pulses with a time bandwidth product (TBP) of the order of unity are achieved, that is, the minimum allowed by the uncertainty principle \cite {TBP}. For FRB, this factor is much greater, for example, for \cite {180404101} high-frequency FRB121102 repetition, component 11H, duration about 0.3 ms, bandwidth about 1 GHz, which corresponds to TBP $ \sim3 \cdot10 ^ 5 $. Thus, the pulse is obviously not as short as can be obtained in laboratory conditions, the duration of the pulse does not correspond to the width of the spectrum and has a different mechanism.

In the works \cite {07094301,171000401,14011795,14121599,14070400,151200529,171100323,160803930,13071628} it was noted that the frequency dependence of the pulse width is $ W \sim \nu ^ \beta $, $ \beta = -4. ..- 4.4 $, which indicates the {\it scatter broadening} phenomenon when the signal propagates through the interstellar medium. The FRBCAT directory gives the scatter indices for 5 signals, all compatible with these values. The strongest scatter broadening manifests itself for 111 MHz low-frequency signals \cite {ATel11932}, for which the pulse width reaches several seconds instead of milliseconds at higher frequencies, which also agrees with an index of about $ -4 $. At high frequencies around 8 GHz, the effect of the interstellar scatter broadening is not as strong, and the observed durations in fractions of milliseconds can be an intrinsic signal characteristic or imparted by the local environment of the FRB source, see \cite {180404101}. In the works \cite {171000401,14011795} it was also noted that for some signals broadening cannot be caused by the interstellar medium and most likely caused by a dense plasma near the FRB source. Both mechanisms are external and also can work for our model, scatter broadening in the interstellar medium and/or dense plasma surrounding the RDM star.

Another characteristic is the burst delay with respect to the time of the asteroid's fall. In \cite {static-rdm} it was noted that a light pulse or an ultrarelativistic particle can dive into the RDM star, reflect from its center and go out, while it will undergo a significant time delay by the clock of the remote observer. The radial light geodesic equation has the form $ dt / dr = \pm \sqrt {B / A} $. Most of the time such a particle spends hanging near the Schwarzschild radius and the directly adjacent inflationary area, where large values of $ B / A $ are reached. The delay for the considered galactic parameters is $ 1.58 $ years for the particle entrance, and the same for the exit. This is an echo of Schwarzschild's behavior, for which the particle from the point of view of a remote observer hangs on the event horizon forever.

In our case, the QG-border is located at the beginning of the inflation area and this result should be corrected. The calculation carried out in Appendix shows that the ultrarelativistic particle spends $ \sim2.5 $ hours on the entry from the ISCO radius $ r = 3r_s $ to $ r_ {QG} $ and the same amount on the exit. To move inside the core, we must accept some assumptions about the behavior of metric coefficients. The ambiguity of doing this can be attributed to the unknown equation of state (EOS). We will consider the metric coefficients frozen on the values on the surface of the core. Integrating the geodesic equation, we get $ t \sim r_s (B_ {QG} / A_ {QG}) ^ {1/2} \sim r_s / \epsilon $, which for the selected parameters gives 3.17 years on entry and the same amount on exit. In this calculation, it was assumed that the shockwave actually reaches the center of the star before it is reflected. If it is reflected earlier from possible structures inside the star, then its exit will also occur earlier. Summarizing, after the asteroid has fallen on the surface of the RDM star, with the assumptions made, the outgoing FRB will follow only after $ \sim6.34 $ years, by the clock of a remote observer.

{\it Repeating bursts.} If several asteroids fall on an RDM star, with a time interval $ dt $, then the interval between individual FRB flashes will also be $ dt $, and the entire sequence will get a delay of $ \sim6.34 $ years. This follows from the stationarity of the solution, radial geodesics have the same shape and are shifted in time by $ dt $ according to the clock of the remote observer. If we consider the details of the process, several shockwaves will be generated that can pass through each other or be reflected in a collision. Several shockwaves will be formed at the exit, accompanied by photon flashes of stimulated radiation. Note that these shockwaves can interact non-trivially with the excited states in the energized volume. In the model, this interaction can have the simplest form if each converging wave excites not all particles present in the volume, but a certain percentage of them, gradually increasing the deposited energy. Also, the diverging wave can stimulate the radiation of not all particles, but some of them, so that successive flashes will have similar energy and intensity. This process is similar to what is happening in the laboratory laser, in which parallel mirrors are installed, so that the light pulse passes through the medium many times to stimulate as many transitions as possible.

Currently, the only one recurring source is registered, with the number FRB121102 for the first event. The FRBCAT catalog lists many recorded repeated flashes with the same coordinates, \cite {180404101} registers 21 flashes over an hour; \cite {180903043} reports 72 more flashes in 5 hours, in the same data, using improved detection algorithm; \cite {181110748} explores the subtle structure of the bursts (sub-bursts). The cause of repeated bursts may be the passage of a massive object through the asteroid field, which was already assumed in \cite {160308207}. If the FRB has the form of a narrow beam, then the sequence of asteroids can come from the same direction, or bombard the source from all sides, covering all possible directions.

In fact, there can be other repeaters. In this regard, we note that the BSA LPI observations \cite {ATel11932} give for FRB160920 and FRB170606 the same angular coordinates, within the beam width of $ 30 '$. The events have very different DM (1767 pc/cm$^3$ and 247 pc/cm$^3$), that is, these are either two sources randomly located on the same line or one source with strongly variable DM. Similarly, \cite {14120342} describes the registration of FRB140514 on the Parkes telescope, whose angular coordinates differ by $ 9 '$ from FRB110220, another observation of Parkes. These events also coincide in the sky with the accuracy of the beam width of $ 15 '$, and have different DM. However, this situation is completely different. FRB140514 was found as a follow-up of FRB110220, in a special search, concentrated in a small neighborhood of the previous event, which increases the likelihood of registering the different sources with close angular distances. While FRB160920 and FRB170606 were detected during the processing of 2012-2018 archived data and, apparently, do not have this kind of correlation. Note that the probability of random coincidence, with an accuracy of $ d \alpha = 30 '$, of two isotropically distributed vectors on the sphere is $ P = \pi d \alpha_ {rad} ^ 2 / (4 \pi) = d \alpha_ {rad} ^ 2 /4\sim1.9\cdot10^{-5}$, extremely small. We also note that \cite {181110748} reports a possible small change in $ \Delta DM \sim1 \mhyph3 $~pc/cm$^3$ over 4 years, for the repeater FRB121102, attributing this to a possible change in the local environment of the source. If FRB160920 and FRB170606 have the same source, then the change in DM that occurred with it indicates extremely large changes in its environment.

Update: Pushchino team published recently the details of their analysis \cite{181210716}, from which it becomes clear that FRB160920 and FRB170606 were detected in a subset, preselected from the daily data, where the same beam of the telescope was pointed to the same small portion of the sky. This makes the detection of different sources on the same line of sight more probable.

{\it Dispersion measure (DM)} characterizes the time delay of the signal, depending on the frequency as $ dt _ {\nu} \sim \nu ^ \gamma $. This delay is usually attributed to different propagation speeds of the electromagnetic signal in the interstellar medium, while the index with a high degree of accuracy is $ \gamma = -2 $. This delay determines the distance to the source and the cosmological redshift factor $ z $, from which the total isotropic energy is extrapolated. However, as \cite {171000401,14011795} note, if scatter broadening occurs on the surrounding source environment, then a significant portion of DM can be also attributed to it. If the environment changes over time, DM also changes. If this is the case, then the many DM-based estimates will be invalidated and should be redone.

{\it Polarization.} FRBs often exhibit a large degree of linear or circular polarization, see FRBCAT, as well as \cite {181009459,180809471} and references therein. This indicates the presence of strong magnetic fields near the source. To describe the polarization of the radiation of RDM stars, a more sophisticated model will be needed, accounting also for the influence of strongly curved space-time in their interior.

{\it Variability} of different parameters between different bursts. We made calculations for the parameters of the central black hole in MW galaxy, while the radiation can come from (quasi) black holes of other sizes. From (\ref {lamout2}) it can be seen that the observed wavelength of the FRB is preserved under scaling $ r_s \to r_s s $, $ \epsilon \to \epsilon s ^ 2 $. The values of $ A_ {QG} $ are also preserved, as well as the values of the energy of incoming particles $ E_N $ and spectral efficiency $ \eta_ {spec} $. Thus, the mentioned FRB characteristics will be the same under
\begin{equation}
\textrm{hypothesis 1:}\quad \epsilon\sim r_s^2. \label{hyp1}
\end{equation}
On the other hand, there are estimates that are not invariant with respect to this scaling. The geometric width of the beam $ D / r_s $ begins to prevail over the wave one $ \lambda / D $ when $ r_s <D ^ 2 / \lambda $, which at $ D = 10 $ km and $ \lambda = 10 $ cm gives $ r_s < 10 ^ 9 $ m. If we take $ r_s <10 ^ 8 $ m with a margin and repeat the calculation (\ref {etas}) for the size of the object falling into the RDM star, at which the beam efficiency can cover the spectral losses, we get $ D = 2.5 \cdot10 ^8 $ m, two diameters of Jupiter. Moreover, $ D> r_s $, the size of the object in this case exceeds the size of the RDM-star, so that the radiation occurs in an isotropic rather than in a radial mode. This makes $ \eta_ {beam} \sim1 $, as a result of which spectral losses can reduce the signal below the registration level. Therefore, in order to ensure high beam efficiency, one should also take
\begin{equation}
\textrm{hypothesis 2:}\quad r_s\gtrsim10^8\textrm{m}. \label{hyp2}
\end{equation}
As a result, the stellar black holes with $ r_s \sim10 ^ 4 $ m are eliminated from the potential sources of FRB. This strongly narrows the set of candidates to the central supermassive black holes, one per galaxy. Note that in \cite {14011795,151200529}, the centers of galaxies also have been the preferred location for FRB sources, to explain the scatter broadening of the pulse. Such scenario automatically leads to the extragalactic origin of FRBs. An exception will be only in the unlikely case that an asteroid falls on the central black hole in our galaxy and the beam of the flash casually points to the Earth. In this case, an extremely powerful intragalactic FRB will be registered.

Generally, there is still a possibility that hypothesis 2 is not fulfilled, for example, if the spectral losses are small, then even the isotropic FRB emission mode can have sufficient power to be registered. In this case, stellar black holes can also be sources of FRB. It is an experimental fact, however, that there are practically no intragalactic FRBs. According to FRBCAT, only one flash of FRB170606 has $ DM \lesssim DM_ {MW} $, on the border. Another flash of FRB010621 also fits into the parameters of the intragalactic one, since its line of sight passes through the galactic center of MW and a large DM is accumulated on it, close to the observed one \cite {frbdis, 14020268}. Also, the spatial distribution of sources is isotropic, and there are no signs of their concentration in the plane of our galaxy \cite {170310173,181110755}. This fact may have a statistical explanation if a single FRB event for some reason has a small probability, which is evenly distributed over stellar black holes in billions of galaxies, then recorded 66 flashes will most likely occur outside our galaxy \cite {181110755}. With a small probability they can occur inside, in which case extremely powerful FRBs will be recorded.

{\it Restrictions from FRB observations to RDM model.} The property (\ref {hyp1}) has interesting consequences for the RDM model. According to (\ref {rhopeff}), at large distances, where $ A \sim1 $, the total flux of dark matter momentum through the $ r $-sphere is $ 4 \pi r ^ 2p _ {\mbox {\it \footnotesize eff}} = \epsilon / 2 $. This flow is divided into two parts, the outgoing flow with an outwardly directed momentum and the incoming flow with an inwardly directed momentum, each part giving the flow contribution $ \epsilon / 4 $ of the same sign. For the considered dark matter of null type (NRDM), the energy is equal to the absolute value of the momentum, so that the outgoing energy flow $ \epsilon / 4 $ is compensated by the incoming $ - \epsilon / 4 $, and the system as a whole turns out to be stationary. The values of $ \epsilon / 4 \sim10 ^ {- 7} $ correspond to the mass flow of $ dm / dt \sim4 \cdot10 ^ {28} $ kg/sec, or 2\% of the solar mass per second, in one direction and the other.

On the surface of the Planck core, the energy density corresponds to the Planck value, also as the momentum flux density, equal to it in geometric units. These values are the same for RDM stars with different parameters, but this is not surprising, since this property is fulfilled by the definition of the Planck core. Namely, the QG-cutoff is performed exactly in the place where the flux density reaches $ l_P ^ {- 2} $. At this point, the total flow through the $r_s$-sphere is proportional to its area and is $ E_0 dN / dt_0 = 4 \pi r_s ^ 2 / l_P ^ 2 $, where $ E_0 $ is the energy of dark matter particles, $ dN / dt_0 $ -- the number of particles per a unit of local time. The total flow we consider is the flow of the momentum in the form of a sum over the incoming/outgoing flows, or the double flow of energy for one flow.

Considering different RDM stars, we assume that locally identical conditions are realized on the surface of the Planck core, so that $ E_0 $ is the same, and $ N $ changes in proportion to the surface area of the core. The energies of dark matter particles inside and outside the RDM stars are related by the redshift factor $ A_ {QG} ^ {1/2} $, given by the formula (\ref {Aqg}). The key point is that (\ref {hyp1}) holds only if the values of this factor are the same for different RDM stars.

Considering the distances $ r> 10r_s $, but much earlier than the outer boundary of the galaxy $ r \ll r_1 $, the coefficient $ A $ is already close enough to 1. At the same time, to obtain the energy and the flow of particles, both $ E_0 $ and $ dN / dt_0 $ are multiplied by the redshift factor $ A_ {QG} ^ {1/2} $, so that the total flow is $ E_1 dN / dt = A_ {QG} E_0 dN / dt_0 = A_ {QG} \cdot4 \pi r_s ^ 2 / l_P ^ 2 = \epsilon / 2 $, according to (\ref {Aqg}). Further variation of flux density, equal to the total flux divided by the area, is a purely geometric effect, $ \sim r ^ {- 2} $. In this case, the pressure of dark matter in geometric units is equal to flux density, $ p = E_1 dN / dt / (4 \pi r ^ 2) $.

We consider RDM stars located in the centers of galaxies and determining the radial distribution of dark matter in them. Further, we assume that all galaxies are surrounded by a reservoir of dark matter with constant pressure and temperature. Somewhere, the galactic pressure is joined to the external pressure $ p_1 = \rho_1 $, this condition determines the outer boundary of the galactic halo $ r_1 $, where the density of dark matter ceases to obey the law $ \sim r ^ {- 2} $. According to our assumption, the temperature of dark matter, that is, the energy of the particles $ E_1 $, is common to all galaxies. Since we also assumed that the energy $ E_0 $ of the particles in the Planck cores of different RDM stars is the same, the factor $ A_ {QG} $ will be the same for all RDM stars. Hence, (\ref {Aqg}) implies (\ref {hyp1}), in the stationary state, the flows regulate themselves so that $ \epsilon \sim r_s ^ 2 $ is satisfied for all RDM stars.

Another interesting aspect is the QG-based explanation of negative masses appearing in the RDM model. It is well known that QG-bounce effectively creates negative energy densities. As explained in \cite {151100633}, when Planck densities are reached, a quantum modification of the equations arises, resulting in effectively negative energy density and repulsive forces, which cause the bounce. In \cite {14016562} this effective density is written as $ \rho_X = \rho \, (1- \rho / \rho_P) $, whence it is immediately seen that exceeding $ \rho> \rho_P $ leads to $ \rho_X <0 $. The stationary state considered in the RDM model can be considered as a frozen QG-bounce, in which a permanent excess of the Planck density leads to a negative effective mass of the core.

This idea can be confirmed by concrete calculation. Suppose the core is incompressible, with a constant density $ \rho $, which corresponds to the effective density $ \rho_X = \rho \, (1- \rho / \rho_P) $. This effective density determines the Misner-Sharp mass: $ M_ {MS} = 4 \pi r_s ^ 3 \rho_X / 3 = r_s (1-B_ {QG} ^ {- 1}) / 2 $. Note that the small positive values of $ 0 <B \ll1 $, which we obtain in the RDM model in the region of inflation, correspond to the large negative $ M_ {MS} $. Substituting here the formula (\ref {ABqg}) from Appendix, we get $ M_ {MS} \sim-4 \pi \epsilon r_s ^ 3 / l_P ^ 2 $, whence $ \rho_X \sim-3 \epsilon / l_P ^ 2 $. If the excess over the Planck density is relatively small, $ \Delta \rho = \rho- \rho_P \ll \rho_P $, from here we get $ \Delta \rho / \rho_P \sim3 \epsilon $. Thus, we have shown that under the influence of the external pressure of dark matter $ \sim \epsilon $, the Planck core slightly increases its density $ \sim3 \epsilon $ so that the resulting anti-gravity force balances this pressure.

Note that the work \cite {151100633} criticizes the Planck star model \cite {14094031,14016562} due to a white hole {\it Eardley's instability} \cite {Eardley, OriPoisson} present in it. On the other hand, in our work \cite {wh-stability}, we showed that this problem can be solved if the core of the Planck star has a negative mass. From our point of view, everyone is right in this dispute, since a slight excess of the Planck density can create negative masses effectively and thereby neutralize the instability.

Interestingly, the Planck length, which initially participated in the calculations, fell out of all the formulas that determine the observable values, such as (\ref {lamout2}), (\ref {lamout4}), (\ref {etas}), (\ref {hyp1}), (\ref {hyp2}). This means that other mechanisms other than QG can give the same answer if they have a core of density $ \rho \sim \lambda_ {X} ^ {- 2} \ll l_P ^ {- 2} $, filled by particles with a Compton wavelength $ \lambda_ {X} \gg l_P $, the mass $ m_X \ll m_P $. In this way, softer mechanisms can be realized, without ultrahigh relativistic factors.

Finally, the energy pumping mechanism for stimulated emission in FRB may be completely different from the inelastic scattering considered here and be similar to the maser mechanism \cite {170306723}, only occur at a different frequency. The main idea of the model considered in this paper is that a gain media for the cosmic laser may be located in the region of strong redshift. An external trigger in a form of a falling asteroid illustrates the possibility of using a strong gravitational field for the energy pumping. There may also be an internal pumping process, which must be energetic enough to overcome the redshift for outgoing photons. For example, a gamma laser operating within a compact massive object with a wavelength on the order of the Compton length of a nucleon, after a wavelength shift of 14 orders of magnitude, can produce an FRB in the decimeter range.

\section{Conclusion}

In this work we have considered RDM-stars as possible sources of FRBs. The QG-cutoff creates a core of Planck density, filled with a gas of Planck mass particles, inside the star. FRBs are generated via the following mechanism. An asteroid falls onto an RDM-star. Large gravitational forces available in the interior accelerate the asteroid to extremely high energies, transforming it to a Lorentz-contracted thin ultrarelativistic shell. The shell falls onto the core and produces a radially converging shockwave, energizing the medium in a conic region of the core. The same shockwave after a bounce from the center, initiates a stimulated emission of energetic photons, releasing the energy, deposited to the medium, in a form of a short pulse of a coherent electromagnetic radiation. Further, strong redshift factor of RDM-star downscales the frequency of this radiation, leading to an observable FRB.

The estimation of an upper frequency, based on the kinematics of deep inelastic reactions between the nucleons composing the asteroid and Planck particles in the core, lead to a formula, involving only the nucleon mass, the gravitational radius and the flux of dark matter in RDM-star. Substituting parameters of the central black hole in the Milky Way galaxy, we obtain the value 0.6GHz for this frequency. Introducing attenuation factors, regulating QG-switch and particle granularity for the asteroid, we obtain the range 0.35-8GHz for the characteristic upper FRB frequency. This estimation fits well with the observable range 0.111-8GHz for the FRBs. 

We also discuss the reconstruction of other FRB parameters in frames of the model, including total energy, energy spectrum, shape of the beam and temporal characteristics of the pulse. FRB observations impose constraints on parameters of RDM-stars used as a model of dark matter halo in spiral galaxies. In particular, a small variability of frequency and other parameters among FRB population implies a relation between the flux of dark matter and the size of the central black hole in RDM model as proportionality $\epsilon\sim r_s^2$. Also, the nature of the negative mass in the center of RDM-star can be clarified as QG-effect, appearing when the core under the pressure of dark matter slightly exceeds Planck density as $\Delta\rho/\rho_P\sim3\epsilon$.

\paragraph*{Acknowledgment.} Thanks to Kira Konich and Kevin Reinartz for proofreading the paper.

\section*{Appendix: RDM model}

The matter distribution, shown on Fig.\ref{f1} left, in spherical coordinates $(t,r,\theta,\phi)$ with a standard metric
\begin{equation}
ds^2=-A(r)dt^2+B(r)dr^2+r^2(d\theta^2+\sin^2\theta\; d\phi^2)\label{stdmetr}
\end{equation}
and a choice of energy-momentum tensor
\begin{equation}
T^{\mu\nu}=\rho(r)(u_+^\mu(r)u_+^\nu(r)+u_-^\mu(r)u_-^\nu(r)),\label{Tmunu}
\end{equation}
corresponding to the density $\rho(r)$ and velocity fields of T-symmetric outgoing/ingoing radial matter flows $u_\pm(r)=(\pm u^t(r),u^r(r),0,0)$, has a general analytic solution of geodesic equations in the form
\begin{eqnarray}
&&4\pi\rho=c_1/\left(r^2u^r\sqrt{AB}\right),\ u^t=c_2/A, \ u^r=\sqrt{c_2^2+c_3A}/\sqrt{AB}.\label{eq_geode}
\end{eqnarray}
Here $c_{1,2}$ are positive constants and $c_3=u_\mu u^\mu$ takes three possible values $c_3=-1,0,1$, dependently on the type of matter (massive, null or tachyonic). Restricting ourselves to $c_3=0$, null radial dark matter (NRDM), the field equations to solve are
\begin{eqnarray}
&&da/dx=-1+e^b+\epsilon\ e^{b-a},\label{eq_dadx}\\
&&db/dx=1-e^b+\epsilon\ e^{b-a},\label{eq_dbdx}
\end{eqnarray}
written in logarithmic variables (\ref{xab}), with $\epsilon=4c_1c_2$ and a starting point
\begin{eqnarray}
&&x_1=\log r_1,\ a_1=0,\ b_1=\epsilon + r_s/r_1.\label{xab1}
\end{eqnarray}
Here $r_s$ is a gravitational radius of RDM-star and $r_1$ is the starting radius of the integration. This system is solved numerically for parameters of MW galaxy in \cite{wh-stability}. The result is presented here on Fig.\ref{f1} right.

Further, the effective density and pressure of dark matter, as energy-momen\-tum tensor components $T_\nu^\mu=\mbox{diag}(-\rho_{\mbox{\it\footnotesize eff}},p_{\mbox{\it\footnotesize eff}},0,0)$ are given by (\ref{rhopeff}). Then, the Planck condition $ \rho _ {\mbox {\it \footnotesize eff}} = l_P ^ {- 2} $ in logarithmic variables has the form $ a = \log (\epsilon l_P ^ 2 / (8 \pi)) - 2x $. For the RDM solution with the MW galaxy parameters in the graph Fig.\ref {f1} on the right, we get the intersection at
\begin{eqnarray}
&&r_{QG}\sim r_s=1.2\cdot10^{10}\textrm{m},\ a_{QG}=-224.591,\ b_{QG}=-195.128, \label{QGab}
\end{eqnarray}
or
\begin{eqnarray}
&&A_{QG}=2.89\cdot10^{-98},\ B_{QG}=1.81\cdot10^{-85}.\label{QGAB}
\end{eqnarray}
Due to the proximity of $ r_ {QG} $ and $ r_s $, for the obtained value of $ A_ {QG} $, the formula (\ref {Aqg}) is valid with high accuracy. For the value of $ b_ {QG} $, the formula $ b_ {QG} = a_ {QG} -2 \log \epsilon $ is also valid with high accuracy, that is
\begin{eqnarray}
&&A_{QG}=(l_P/r_s)^2/(8\pi)\cdot\epsilon,\ B_{QG}=(l_P/r_s)^2/(8\pi)/\epsilon.\label{ABqg}
\end{eqnarray}
Note that according to estimates of \cite {static-rdm}, in the inflation area, $ b = a + c_9 $, $ c_9 \sim-2 \log \epsilon $ holds in the initial interval, in the leading order in $ \epsilon $. This relationship also holds, not only in order, but with high accuracy, in our numerical experiment with selected model constants. Another relation from \cite {static-rdm}, $ b_2 = -a_2-c_8 $, describes a small deepening of the maximum of $ b_2 $ compared to Schwarzschild mode $ b = -a $, which, apparently, is a local effect, not propagating in the adjacent asymptotics.

Another subtle effect is the small difference between the point $ r_2 $ and the nominal Schwarzschild radius $ r_s $ noted in \cite {static-rdm, wh-stability}. While the nominal radius is used in the formulation of the initial value $ b_1 $, at large distances, found by asymptotic formulas, the point $ r_2 $ corresponds to the maximum of the $ b $ -function located in the region of strong gravitational fields and found with an accurate account for all effects in the result of numerical integration. For the used parameters, $ r_2 $ is lower than $ r_s $ by 10\%. Note that the value of $ r_s $ itself was reconstructed from the orbits of S-stars in \cite {Ghez} with an accuracy of about 10\%. Therefore, this difference has no practical value, but we are interested in the principal question of the model calibration.

As noted in \cite {static-rdm}, other features of the gravitational field, such as the innermost stable circular orbit (ISCO), correspond to the value of $ r_2 $, rather than the nominal $ r_s $, which means that in Schwarzschild mode, the $ r_2 $, not $ r_s $, plays the role of a gravitational radius. We also calculated the orbital velocity using exact formulas from \cite {static-rdm}, $ v ^ 2 = da / dx / 2 $, for the position of S-stars $ r \sim2 \cdot10 ^ {13} $ m ... $ 3 \cdot10 ^ {14} $ m and found that the correspondence with the Newtonian approximation holds better for $ v ^ 2 = r_2 / (2r) $ than for $ v ^ 2 = r_s / (2r) $.

Because of this, we recalibrated the model, selecting the nominal value of $ r_ {s, nom} $ so as to get $ r_2 = r_ {s, exp} $. The required nominal value turned out to be $ r_ {s, nom} = 1.32 \cdot10 ^ {10} $ m. With such a nominal value, integration was performed, the results of which were presented in Fig.\ref {f1} on the right, and for which the cutoff point was obtained (\ref {QGab}). In the formulas (\ref {Aqg}), (\ref {ABqg}), the experimental value of $ r_s $ is used, which coincides with $ r_2 $ as a result of recalibration.

We also compared the integration results for NRDM parameters with the tabular \cite {wh-stability} results without recalibration. It turned out that the $ a, b $ parameters at the critical points practically do not change, and the entire form slides down in $ x $ by $ 0.10 $-$ 0.11 $, which corresponds to scaling of $ r $ by $ 10 $-$ 11 $\%. However, this change is not the scaling precisely. Although the equations to be solved are autonomous, and the scaling of the solution belongs to the exact symmetry group, the starting value of $ b_1 $, and only it, contains dependence on $ r_s $. However, this contribution is small compared to the dark matter term contained in $ b_1 $, which, apparently, explains the smallness of the variation of $a,b$-functions under the described modification of the solution.

To determine the time delays, we have integrated the radial light geodesic equation on the found solution. In logarithmic variables, this equation has the form:
\begin{equation}
dt/dx=\pm e^{x+(b-a)/2}.\label{eq_geode2}
\end{equation}
A similar to \cite {static-rdm} numerical integration of this equation between the critical points $ x_2 $ and $ x_3 $ for the selected galactic parameters gives the time interval of $ 1.58 $ years. More relevant for our formulation of the problem is this integral between $ r = 3r_s $ and $ r = r_ {QG} $, which gives $ \Delta t = 2.5 $ hours. Further, inside the Planck core, we considered the metric coefficients as frozen and for the dive time from $ r = r_ {QG} $ to $ r = 0 $, obtain the interval $ t \sim r_s (B_ {QG} / A_ {QG}) ^ {1 /2}\sim3.17$~years.

Due to the fact that for the RDM model in the region of inflation, the metric coefficients are extremely small, a peculiar phenomenon of gravitational beaming arises. The displacement of a light particle in the radial direction $ dt = (B / A) ^ {1/2} dr \sim dr / \epsilon $ requires much less time by the clock of the remote observer than the comparable $ dr \sim rd \theta $ shift in the tangential direction $ dt = A ^ {- 1/2} rd \theta $. In particular, when choosing the parameters (\ref {QGAB}), to move the particle in the radial direction by $ dr = 1 $ m, 8ms is required, while to move the particle in the tangential direction by $ rd \theta = 1 $ m, one needs $ 2 \cdot10 ^ {40} $ s, which is 23 orders of magnitude greater than the age of the universe. Practically, this means that the particles do not move in the tangential direction, from the point of view of the remote observer, as long as the particles are located in the region of small values of the metric coefficient $ A $.

In particular, the rotation of the core is such tangential displacement. Of course, if there is a rotation, the problem loses its spherical symmetry and instead much more complicated problem of the Kerr-Newman type should be solved. Here we make only an estimate, assuming that the surface of the core rotates at a speed close to the speed of light, then its rotation by $ 1 '' $ with the selected parameters would require $ \sim10 ^ {45} $ s, which is 27 orders of magnitude greater than the age of the universe. In practice, this means that, according to the clock of the external observer, the rotation of the Planck core stops, at least for the spherically symmetric RDM solution considered here. 

For $ \lambda \sim0.1 $ m, $ D \sim10 ^ 4 $ m, the angular width of the beam is $ \lambda / D \sim10 ^ {- 5} $, and if we assume the distance to the source $> 10 ^ 5 $ light years, the linear transverse size of the beam at that distance will be $> 1 $ light year. With the possible transverse velocity of the observer relative to the source $ \sim300 $~km/s~\cite {180706658},\linebreak the observer will cross the beam $> 1000 $ years. Practically, if an FRB beam in the past pointed to the Earth, then it will continue to point to it in further observations, and repetitions of the process in the same region of the Planck~core will be recorded as the repeating bursts.

\end{document}